# Nonlinear Optical Limiting with Hybrid Nanostructures of $NiCo_2O_4$ and Multiwall Carbon Nanotubes

Ankit Sharma[1†], Pritam Khan[2*†], Mansi Pathak[3], Chandra Sekhar Rout[3], and K. V. Adarsh[1*]

[1]Department of Physics, Indian Institute of Science Education and Research Bhopal, India

[2] Department of Physics and Bernal Institute, University of Limerick, Limerick V94 T9PX, Ireland

[3] Centre for Nano and Materials Science, Jain University, Ramanagaram, Bangalore, 562112, India

**Abstract:** Nonlinear optical (NLO) response in terms of reverse saturable absorption (RSA) has been exploited extensively for optical limiting. Here, we experimentally demonstrate that flower-like hybrid nanostructures of $NiCo_2O_4$ and Multiwall Carbon Nanotubes (NCO@MWCNT) exhibit a strong non-linearity in their absorption, specifically an excited state absorption (ESA) induced RSA, when exposed to nanosecond laser pulses. We obtain strong nonlinear absorption coefficient (β) and nonlinear refractive index ($n_2$) in hybrid NCO@MWCNT, the values that are 2-times and 2-orders of magnitude higher, respectively compared to NCO or MWCNT alone. This offers straightforward application in optical limiting with optical limiting ($F_{OL}$) and optical onset ($F_{ON}$) threshold ca. 2-10 times lower than benchmark NLO materials, e.g., graphene, family of 2-D transition metal dichalcogenides (TMDC) materials and recently established NCO. Notably, for femtosecond pumping, NLO response of NCO@MWCNT is dominated by saturable absorption (SA) with a week contribution from two photon absorption (TPA), arising respectively from MWCNT and NCO. A remarkable sign reversal along with a larger amplitude is obtained for $n_2$ thanks to the charge transfer from MWCNT to NCO.

*Authors to whom correspondence should be addressed: pritam.iiserb@gmail.com & adarsh@iiserb.ac.in

† These authors contributed equally to this work.

# Introduction

Materials exhibiting strong nonlinear absorption and refraction form the building blocks of numerous photonics devices in optics and optoelectronics. In general, NLO materials can be classified in two types based on (i) SA, i.e., increased transmittance at high intensity and (ii) RSA, i.e., decreased transmittance or high absorption at high intensity [1,2]. While SA find applications in mode locking and Q-switching [3,4], RSA is useful for developing optical limiter [5] and optical sensor [6] etc. Optical limiting in particular is very important since it can be used to protect human eyes and sensors from potential optical damage. They do so by allowing reduced transmission at high laser intensity and keeping high transmission at lower intensity values [5, 7, 8]. There have been significant developments over past few decades for the designing of optical limiters with large NLO responses, particularly with low dimensional carbon-based materials in comparison to their bulk counterparts [9]. This includes single-walled and multi-walled carbon nanotubes (CNTs) i.e., SWCNT and MWCNT [8,10-11], fullerenes [12], carbon black suspensions [13], porphyrins [14] etc.

For over two decades now, CNTs are established as benchmark NLO material for optical limiting. A plethora of research have been therefore conducted and showed that the NLO response is improved significantly by fabricating covalent/noncovalent composite or hybrid structures of both SWCNT and MWCNTs with secondary materials having complementary NLO properties [15-18]. Since the formation of a composite structure requires the dispersion of CNTs into host matrix, MWCNTs are preferred over SWCNTs because MWCNTs disperse easily and effectively [19]. Apart from that effect of small bandgap (as low as ~ 2kT) and intershell coupling by van der Waals interaction allows nonradiative transition, thereby making MWCNTs as good optically active material [20].

In our recent work, we have shown that spinel NCO shows strong third order NLO response with obvious application in optical limiting thanks to ESA [21]. This observation prompted us to investigate the optical limiting behavior of noncovalently connected hybrids of NCO and MWCNT, thus NCO@MWCNT. In our earlier work, we have synthesized NCO@MWCNT and demonstrated that it can be explored as high-performance supercapacitor electrodes [22]. Therefore, observation of NLO in NCO@MWCNT will right away open up encouraging multimodal electro-optical applications.

We demonstrate here that when MWCNT is functionalized with NCO, NCO@MWCNT exhibits strong ESA that provides straightforward application in optical limiting. The two critical parameters for optical limiting, i.e., optical limiting ($F_{OL}$) and optical onset ($F_{ON}$) threshold achieved here with nanosecond laser pulses is 2-10 times lower than NCO, graphene, and other family of 2-D transition metal dichalcogenides (TMDC) materials, thus sets up new performance milestone. Optical limiting application is supported from open-aperture and closed-aperture Z-scan results that indicate that $\beta$ and $n_2$ is 2-times and 2-order magnitude higher in hybrid NCO@MWCNT, respectively compared to NCO or MWCNT alone. We believe that the optical limiting effect in hybrid NCO@MWCNT arises from charge transfer mechanism.

## II. SAMPLE PREPARATION AND OPTICAL CHARACTERIZATION

### A. Synthesis and morphological characterization

We synthesize NCO and NCO@MWCNT by facile hydrothermal method followed by high temperature annealing in air as described in our recent work [22]. The morphologies of the synthesized products were determined by field emission scanning electron microscopy (FESEM,

JEOL JSM-7100F, JEOL Ltd., Singapore) as shown in Figure 1(a)-(c). Flower-like nanoneedle nanostructure is observed for NCO (Figure 1(a)). Clear nanotube like structure in Figure 1(b) confirms the high-quality synthesis of MWCNT. Finally in Figure 1(c), we observed flower-like NCO is decorated on MWCNT in the hybrid system.

### B. Linear optical response

Figure 1(d) shows the optical absorption spectra of all three samples. Tauc plot of NCO exhibit the presence of two bandgaps at 3.24 and 1.67 eV, arising because of the different optically active high and low spin state of Co ion in NCO as described in detail in our recent work [21]. MWCNT shows the broad absorption spectra spreading over UV-Vis and NIR region with a weak shoulder peak around ~280 nm (4.43 eV) in line with several previous observations [23, 24]. For the hybrid system of NCO@MWCNT, the peak shifted to ~ 285 nm (~ 4.36 eV) which we presume to be associated with the band alignment during charge transfer mechanism.

## III. RESULTS AND DISCUSSION

### A. Non-linear optical response with nanosecond pumping

We use the standard open aperture (OA) and closed-aperture (CA) Z-scan techniques to quantify the third-order optical nonlinearities by extracting β and $n_2$, respectively in our samples. For this experiment, we used 7 ns, 532 nm laser pulses generated from the second harmonic of Nd-YAG laser. A low repetition rate of 10 Hz is used to minimize the heating effect and photodamage. The Rayleigh length ($Z_R$) and beam waist ($\omega_0$) in our experiment were ~ 3.7 mm and ~ 25 μm, respectively.

Figure 2 (a) shows the open aperture (OA) Z-scan traces of NCO (red), MWCNT (green), and hybrid NCO@MWCNT (blue), respectively when excited with 532 nm, ns Gaussian laser pulses at 0.49 GW/cm$^2$. While we observed a reverse saturable absorption (RSA) with a weak saturable absorption (SA) shoulder in NCO, MWCNT and NCO@MWCNT exhibit pure SA and RSA characteristics respectively (Figure 2(a)). In our very recent work [21], we showed that RSA in NCO originates from excited-state absorption (ESA) in a two-step sequential process: (i) inter-band transition from the valence band (VB) to low spin state (Co 3d-$t_{2g}$ or Ni 3d-$t_{2g}$) in the conduction band (CB) and (ii) intra-band transition from an already exciting low spin state to high spin state (Co 3d-$e_g$ or Ni 3d-$e_g$) within the CB. By calculating the cross-section for ground ($\sigma_{gs}$), excited-state ($\sigma_{ex}$) using following equations [25, 26]:

$$\sigma_{gs} = \frac{-\log T_0}{NL} \qquad \& \qquad \sigma_{ex} = \frac{-\log T_{max}}{NL} \tag{1}$$

where, $T_0$, $T_{max}$, N, and L are linear transmission, saturated transmission at high intensity, ground state carrier density (sample concentration x Avogadro's number), and thickness of sample respectively, we found for NCO that $\sigma_{gs}$ = 4.63 x $10^{-19}$ cm$^2$, and $\sigma_{ex}$ = 8.33 x $10^{-19}$ cm$^2$. The ratio $\sigma_{ex}/\sigma_{gs}$=1.8 (greater than unity) confirms that RSA observed in NCO is associated with ESA. Moving onto MWCNT, the SA can be explained by assuming at such higher intensity, ground state carriers are depleted, the excited state becomes almost occupied while the Pauli exclusion principle prevents further absorption of photons. Likewise, optical transitions are reduced significantly to give rise to SA with increased light transmission. This explanation is supported by our calculations that shows $\sigma_{ex}$ = 2.00 x $10^{-20}$ cm$^2$ is lower than $\sigma_{gs}$ = 6.12 x $10^{-20}$ cm$^2$, ($\sigma_{ex}/\sigma_{gs}$=0.33) and further establishes MWCNT as an ideal slow (since ns pulse-width is relatively longer) saturable absorber. For the hybrid system of NCO@MWCNT, strong ESA in NCO overcomes the SA in MWCNT to produce an overall RSA behavior mediated by ESA (blue curve, Figure 2(a)).

The observation of ESA is further supported by the calculations which showed that $\sigma_{ex}$ = 4.04 x $10^{-19}$ $cm^2$ and $\sigma_{gs}$ = 1.26 x $10^{-19}$ $cm^2$, thus the ratio ($\sigma_{ex}/\sigma_{gs}$=3.20) is greater than unity. Since the energy of excitation 2.33 eV (532 nm) is approximately half of the bandgap energy of the NCO@MWCNT, i.e., 4.36 eV (285 nm), ESA is presumably driven by the two-photon process. The observed results can be intuitively explained from Figure 2(b). Since the work functions of MWCNT (4.7 eV) and NCO (5.5 eV) are different, when the conductive MWCNT and p-type semiconductor NCO hybridize, electron transfer takes place from MWCNT to NCO until there is only a single common Fermi level. Likewise, Meanwhile, a subsequent electric field is developed between them which induces an upward band bending of 0.8 eV ($\Phi_{NCO}$-$\Phi_{MWCNT}$) towards MWCNT. Optical absorption spectra also support the band alignment in which the low spin state of NCO is completely suppressed in the hybrid system and only the high spin state contributes to optical absorption as shown in Figure 1(d).

To quantify the observed nonlinearity, we used the well-established Z-scan theory [27,28] to determine the intensity dependence of saturation intensity ($I_s$) and nonlinear absorption coefficient ($\beta$). The details can also be found in our recent works [21,29,30]. Experimentally fitted data reveals that in MWCNT, $I_s$ values are 8.04 ± 1.2 and 8.13 ± 0.90 GW/$cm^2$ for peak intensities of 0.36 to 0.49 GW/$cm^2$, respectively. Since $I_s$ remains nearly constant, we envisage that MWCNT is highly advantageous for fabricating nanophotonic devices such as passive mode lockers. Remarkably, we found that $\beta$ increases from ca. 20 to 55 cm/GW in NCO and from 13 to 100 cm/GW in NCO@MWCNT, when the intensity is raised from 0.36 to 0.49 GW/$cm^2$, the observation that is consistent with the theory of ESA. Clearly, we achieve a two-factor enhancement in $\beta$ in the hybrid system compared to pristine NCO at higher intensities. At this point, it is important to note that the observed nonlinearity in NCO@MWCNT not only surpasses NCO but is also found to be the highest among the family of transition metal oxide families (Table T1, Supplemental Material (SM)).

Apart from the β, another important parameter for consideration is the nonlinear refractive index ($n_2$). As opposed to OA Z-scan traces that provide β, CA Z-scan measurements are performed to determine $n_2$ at 0.49 GW/cm$^2$ as shown in Figure 2(c). Notably, to discard the effect of nonlinear absorption, we divided the CA data by OA in Figure 2(c). The change in $n_2$ can take place either from free or bound carrier nonlinearity [31]. During longer nanosecond laser excitation, free carriers determine the variation in $n_2$ by masking the effect of fast-bound carriers. It can be seen from Figure 2(c) that both NCO and MWCNT exhibit a valley-peak (V-P) structure. Such pre-focal transmission minimum followed by a post-focal maximum is consistent with the self-focusing effect that gives rise to positive $n_2$. Our calculations (details can be found in our earlier work [21,32]) show that for NCO and MWCNT, $n_2$ is found to be 1.80 x 10$^{-4}$ and 5.36 x 10$^{-3}$ cm$^2$/GW, respectively. The larger positive $n_2$ obtained in MWCNT compared to NCO can be intuitively explained by free carrier induced SA [33] in MWCNT which is also evinced from the larger post-focal transmission change compared to NCO (Figure 2(c)). Remarkably, for the hybrid NCO@MWCNT system, we observed a sign reversal, i.e., negative $n_2$ with a larger change in normalized transmission, thus very high $n_2$ of ca. -1.18 x 10$^{-2}$ cm$^2$/GW. This means CA Z-scan trace exhibits peak-valley (P-V) structure, i.e., pre-focal transmission maximum and post-focal maximum that results in self-defocusing effect from bound carriers. This is because a strongly coupled system provides a way to sink all carriers by the charge transfer mechanism, which eventually formed bound carriers. To account for the thermal contribution, we calculate the axial peak-valley difference ($|\Delta Z|$), all of which are found to be less than 1.7 $Z_R$ (1.66, 1.55, and 0.74 $Z_R$ for NCO, MWCNT, and NCO@MWCNT, respectively) where $Z_R$ is Rayleigh parameter. This indicates that the observed nonlinear effect is a third-order process with an insignificant thermal contribution.

The high β and $n_2$ obtained for NCO@MWCNT immediately call for application. As an example, it is well known that high β scales linearly with the optical limiting capability [5, 7]. Since two-factor higher β is obtained in NCO@MWCNT, we aim to establish the hybrid system as a superior optical limiting device. As a reference sample, we choose here NCO which is already proven to be a good optical limiter as shown in our very recent work [21]. In this regard, Figure 3(a) and (b) present the variation of output intensity ($I_{out}$) and normalized transmittance against input intensity ($I_{in}$) to demonstrate the suitability of NCO@MWCNT as an optical limiter as well as compare its performance with pristine NCO. The performance of an optical limiter is characterized by three factors: (i) limiting differential transmittance ($T_d = dI_{out}/dI_{in}$) at higher intensity, (ii) onset threshold intensity $F_{ON}$ (input intensity at which the normalized transmittance deviated from linearity), and (iii) optical limiting threshold intensity $F_{OL}$ (input intensity at which normalized transmittance drops below 50%). Notably, the lower the value of $T_d$, $F_{ON}$ and $F_{OL}$, the better is the performance of the optical limiter. Our results indicate that for both the pristine and hybrid system (Figure 3(a)), $I_{out}$ scales linearly with $I_{in}$ at lower intensities, however, deviates from linearity at higher intensities. A quantitative comparison indicates that at 0.64 GW/cm$^2$, $T_d$ is 0.27 for NCO, while for hybrid system $T_d$ gives the lower value of 0.20 at 0.49 GW/cm$^2$. Such observation indicates that NCO@MWCNT attenuates the higher intensity much more efficiently compared to NCO, thus established as a superior optical limiter. To demonstrate device application further, we calculate and compare two other critical parameters $F_{OL}$ and $F_{ON}$ from Figure 3(b). For NCO, $F_{ON}$ and $F_{OL}$ are found to be 0.04 GW/cm$^2$ and 0.52 GW/cm$^2$, while remarkably for the hybrid NCO@MWCNT system, we obtained significantly lower value in $F_{ON=}$ 0.02 GW/cm$^2$ (2 times lower) and $F_{OL}$=0.28 GW/cm$^2$ (2 times lower). Therefore, we establish a hybrid NCO@MWCNT system as an efficient novel nonlinear optical material having optical limiting capability surpassing NCO as well as the well-known benchmark $WS_2$, $MoS_2$, $WSe_2$, and $MoSe_2$ nanosheets, graphene oxide nanosheets, CdS nanoparticles etc. (Table 2, SM).

## B. Non-linear optical response with femtosecond pumping

In this experiment, we employed 120 femtosecond, 800 nm laser pulses generated from Ti-sapphire to excite our sample. The repetitions rate of the incident laser was 1 kHz. For ultrafast pumping, the $Z_R$ and $\omega_0$ were ~ 12 mm and ~ 55 μm, respectively. Figure 4(a) depicts the OA Z-scan traces of the pristine and hybrid systems when excited with femtosecond laser pulses. While pristine NCO and MWCNT exhibit pure RSA and SA respectively, the hybrid NCO@MWCNT trace has contributions from both RSA and SA. For NCO, RSA is characterized by two-photon absorption (TPA) since the excitation energy of 1.55 eV (800 nm) is approximately half the energy (3.24 eV) required for inter-band transition between VB to high spin state in CB [21]. For MWCNT, SA can be explained by assuming that at such higher intensity with fs laser, ground state carriers are depleted, the excited state becomes almost occupied while the Pauli exclusion principle prevents further absorption of photons. Likewise, optical transitions are reduced significantly to give rise to fast SA (fs pulse-width is shorter) with increased light transmission. On the other hand, for the hybrid system, we find that because of band alignment and the common fermi level of MWCNT and NCO, we observed SA and TPA merged together (blue curve).

To determine $n_2$ we performed CA Z-scan measurements with fs laser at 78.9 GW/cm$^2$ as shown in Figure 4(b). Notably, like nanosecond excitation we divided CA by OA to discard the contribution from nonlinear absorption. Remarkably, we could observe a sign reversal, i.e., positive $n_2$ for hybrid NCO@MWCNT (V-P structure) compared to negative $n_2$ of pristine NCO and MWCNT (P-V structure). In fact, the sign reversal in $n_2$ is universal for the hybrid system, irrespective of the lasers used except the fact that it takes place with opposite signature and trend (Figure 2(c) and 4(b)). Our calculations show that $n_2$ is found to be −7.8 x $10^{-8}$, −2.87 x $10^{-4}$, and +1.66 x $10^{-5}$ $cm^2$/GW for NCO, MWCNT, and hybrid system, respectively. Negative $n_2$ in NCO originates from the combination bound carriers and weak thermal contribution, since sub-bandgap fs excitation does not produce any free carriers and since ($|\Delta Z|$= 1.9 $Z_R$) is higher than 1.7 $Z_R$. For MWCNT, $n_2$ is inherent to the SA effect [34] and consistent with the observed effects in CNT of different structures, i.e., single-wall CNT [35], double-wall CNT [36] and other MWCNTs [37]. Apart from that, $|\Delta Z|$= 1.8 $Z_R$ conforms that thermal effect is insignificant in MWCNT. We propose that for hybrid system electron transfer from MWCNT to NCO provide additional free carriers that induced positive refractive index with three orders of higher magnitude as compared to NCO. Also, for NCO@MWCNT we can completely discard the thermal effect since $|\Delta Z|$= 1.8 $Z_R$ <1.7 $Z_R$.

## Conclusions

In summary, we demonstrated an unusually efficient nonlinear optical-limiting behavior in hybrid NCO@NWCNT, i.e., when MWCNT is functionalized with NCO with complementary NLO responses. As a result, practical liquid cell based optical limiters can now be designed with performances exceeding those benchmark NLO materials, e.g., graphene, family of 2-D transition metal dichalcogenides (TMDC) materials by a factor of ca. 5–10. We demonstrated that excited state absorption and charge transfer mechanism contributes to high β and $n_2$ in NCO@MWCNT.

Since NCO@NWCNT is already established as a high-performance supercapacitor, the observation of strong NLO will readily encourage other electro-optic applications.

## Acknowledgement

DAE BRNS (37(3)/14/26/2016-BRNS/37245); Science and Engineering Research Board (CRG/2019/002808); FIST Project for Department of Physics. A. S. gratefully acknowledge the UGC-CSIR for financial assistance.

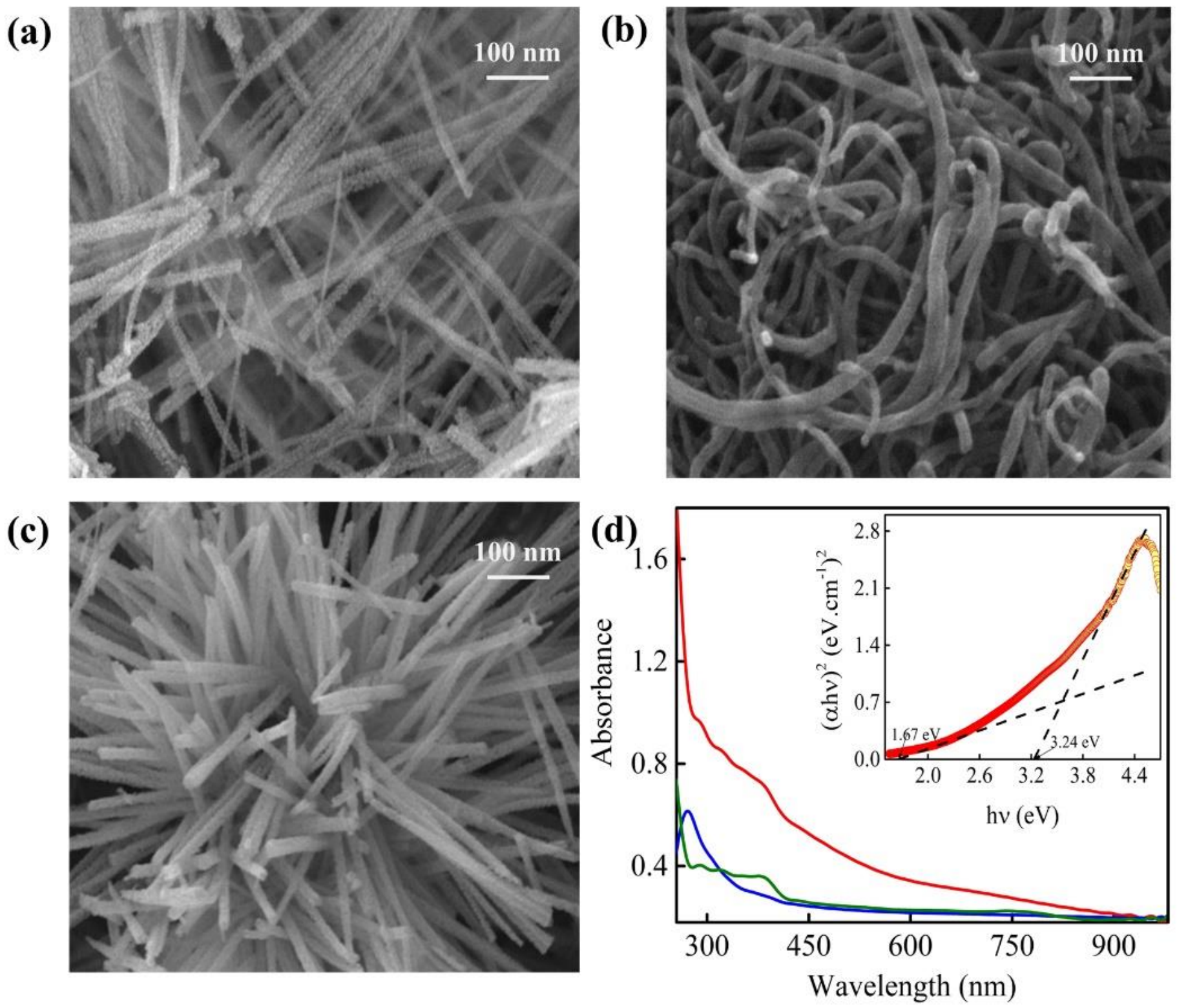

**Figure 1**. FESEM images of (a) NCO, (b) MWCNT, and (c) NCO@MWCNT. The scalebars are shown in the insets of each figure. (d) optical absorption spectra of the samples. (NCO: Red, MWCNT: Green, and NCO@MWCNT: Blue, in color with hollow circle). Inset shows the Tauc plot used to calculate the bandgap of NCO.

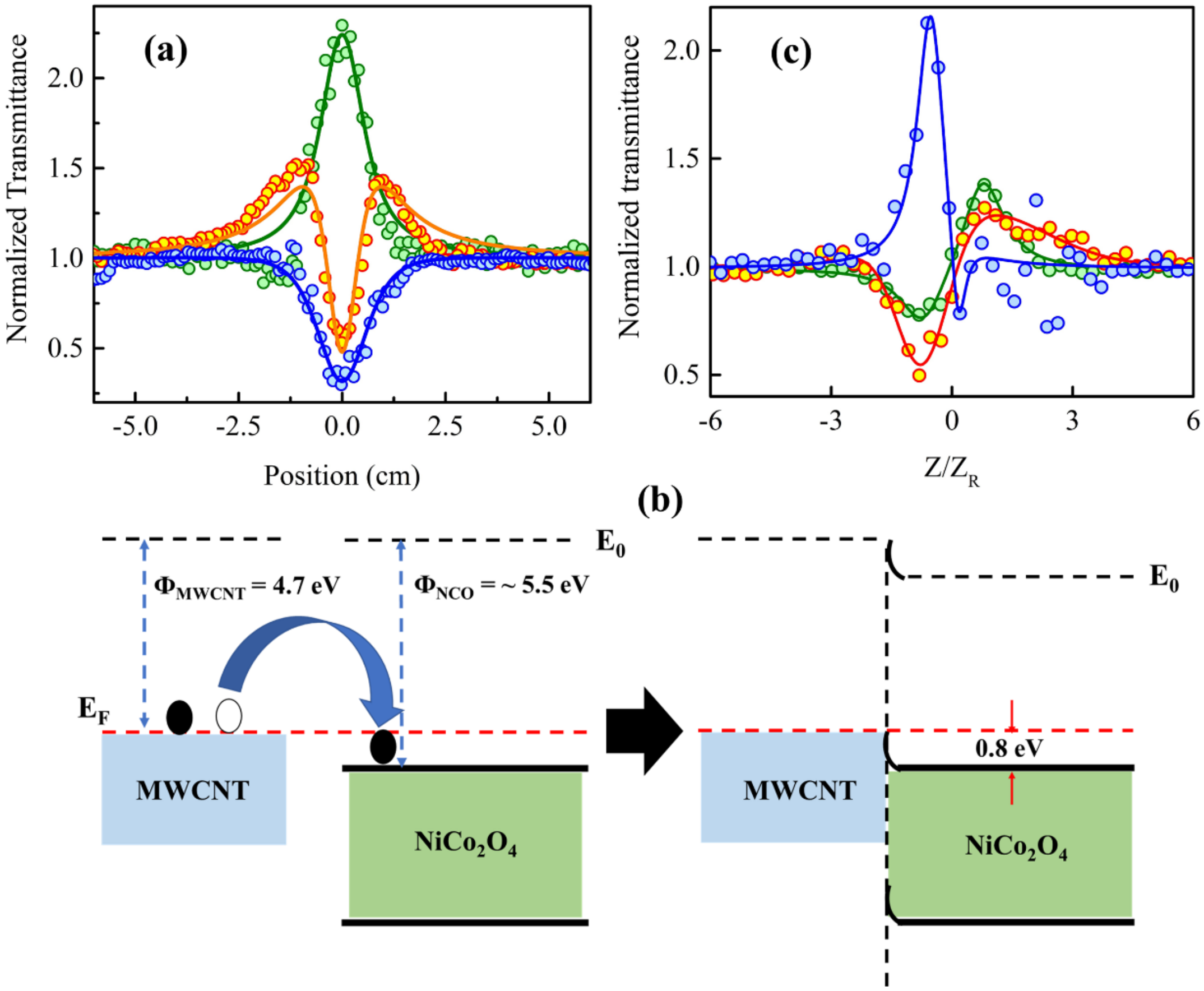

**Figure 2**. (a) OA Z-scan traces of represents of the samples for 532 nm nanosecond laser excitation. (NCO: Red, MWCNT: Green, and NCO@MWCNT: Blue, in color with hollow circles). The circles represent the experimentally observed values and the solid lines imply theoretical fitting; (b) Schematics of band diagram for NCO, MWCNT, and NCO@MWCNT hybrid to describe the charge transfer mechanism. Here, $E_0$ is vacuum level, $E_F$ is the Fermi energy level, Φ is the work function and (c) CA Z-scan traces with nanosecond laser showing the transition from positive to negative $n_2$ with enhanced magnitude in the hybrid system. (NCO: Red, MWCNT: Green, and NCO@MWCNT: Blue, in color with hollow circles).

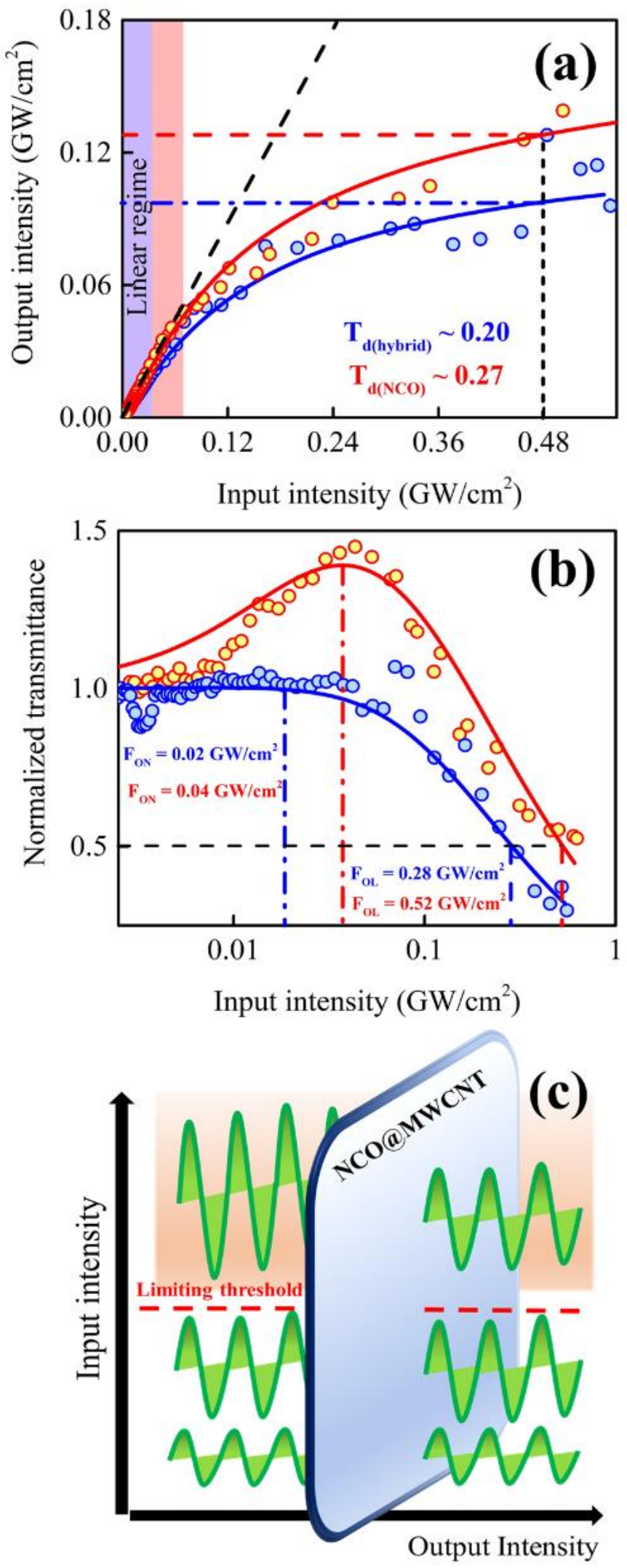

**Figure 3**. (a) Output intensity as a function of input intensity for 532 nm nanosecond pulse excitation. The black dashed and colored solid lines in the figure show the linear transmittance (T) and the theoretical fitting, respectively. (b) Normalized transmittance as a function of input intensity. (c) Schematic diagram demonstrating the optical limiting application with hybrid NCO@MWCNT.

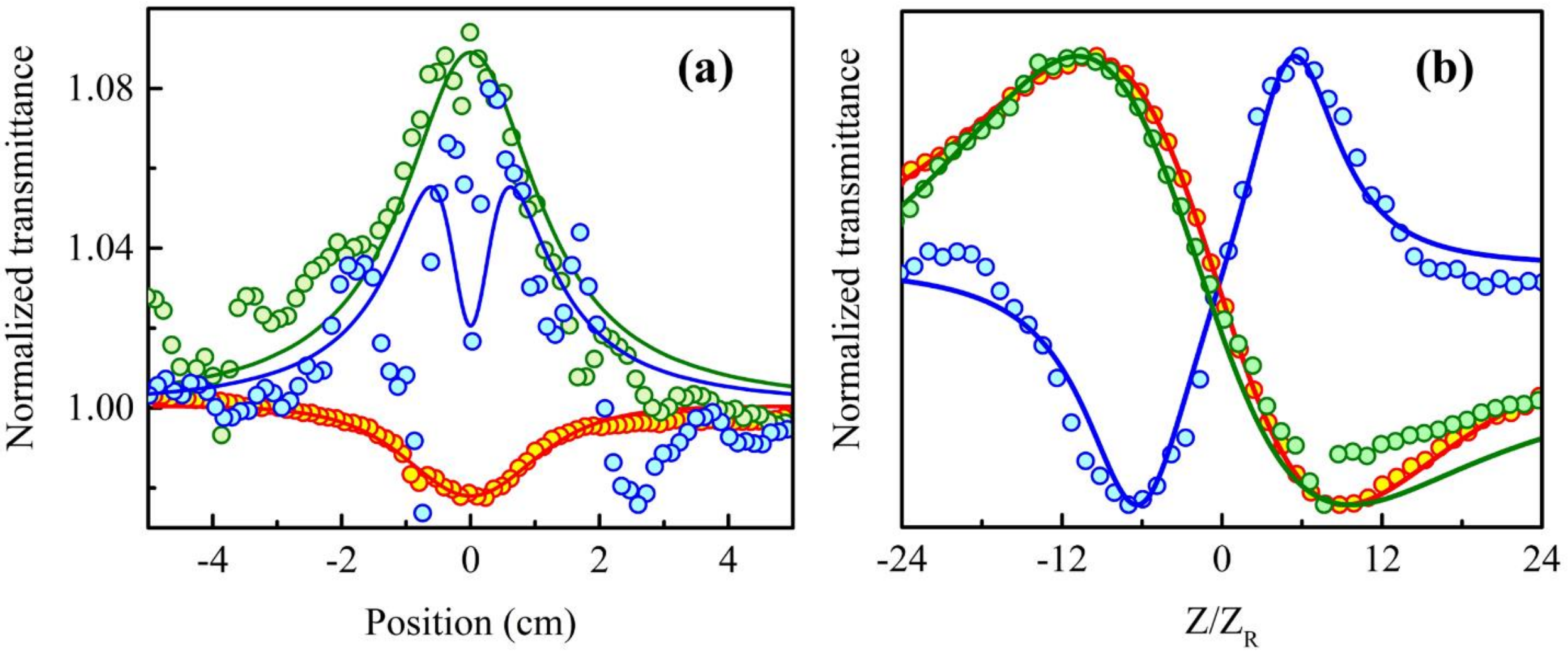

**Figure 4**. OA Z-scan traces of the samples with 800 nm, femtosecond laser pumping. at 63 GW/cm$^2$; (b) CA Z-scan traces indicate the sign reversal of $n_2$ in the hybrid system. (NCO: Red, MWCNT: Green, and NCO@MWCNT: Blue, in color with hollow circles).

# Supplemental Material

## Nonlinear Optical Limiting with Hybrid Nanostructures of $NiCo_2O_4$ and Multiwall Carbon Nanotubes

Ankit Sharma[1†], Pritam Khan[2*†], Mansi Pathak[3], Chandra Sekhar Rout[3], and K. V. Adarsh[1*]

[1]Department of Physics, Indian Institute of Science Education and Research Bhopal, India

[2] Department of Physics and Bernal Institute, University of Limerick, Limerick V94 T9PX, Ireland

[3] Centre for Nano and Materials Science, Jain University, Ramanagaram, Bangalore, 562112, India

**Table T1:** Comparison of nonlinear absorption coefficient (β) and nonlinear refractive index ($n_2$) with previously reported samples when excited with 532 nm nanosecond laser.

| **Sample** | **Pulse duration** | **β (cm/GW)** | **$n_2$ (cm²/GW)** | **Ref.** |
|---|---|---|---|---|
| $NiCo_2O_4$@MWCNT | 7 ns | 100 | $-1.18 \times 10^{-2}$ | Present work |
| $NiCo_2O_4$ | 7 ns | 54.9 | $1.8 \times 10^{-4}$ | Ref. [1] |
| $ZnCo_2O_4$ | 7 ns | 27.4 | --- | Ref. [2] |
| ZnO | 5 ns | 7.6 | --- | Ref. [3] |
| $Fe_2O_3$ | 15 ns | 1.0 | --- | Ref. [4] |
| $Cr_2O_3$ | 4 ns | 3.17 | --- | Ref. [5] |

**Table T2:** Optical limiting parameters for different materials system at 532 nm nanosecond laser excitation.

| | **532 nm** | | |
|---|---|---|---|
| **Sample** | **$F_{ON}$ (J/cm$^2$)** | **$F_{OL}$ (J/cm$^2$)** | **References** |
| **NCO@MWCNT** | 0.14 | 1.96 | Present work |
| **$MoS_2$** | 1.52 | 11.16 | Ref. [6] |
| **$MoSe_2$** | 1.47 | 7.30 | Ref. [6] |
| **$WS_2$** | 1.24 | 9.35 | Ref. [6] |
| **$WSe_2$** | 0.99 | 7.20 | Ref. [6] |
| **Graphene** | 0.44 | 15.15 | Ref. [6] |
| **NCO** | 0.28 | 3.64 | Ref. [6] |
| **CdS nanoparticles** | 0.30 | 2.55 | Ref. [7] |
| **Graphene Oxide** | 0.19 | 1.19 | Ref. [8] |